\documentclass[10pt,prl,showpacs,twocolumn,superscriptaddress]{revtex4-1}

\pdfoutput=1

\usepackage{amsmath}
\usepackage{graphicx} 
\usepackage{dcolumn}  
\usepackage{bm}       

\graphicspath{{./figures/}}

\pacs{05.30.Jp, 03.75.Hh, 67.85.Bc}
\begin{document}

\author{T.~Betz }
\affiliation{Atominstitut, TU-Wien, 1020 Vienna, Austria}
  
  \author{S.~Manz}
 \affiliation{Atominstitut, TU-Wien, 1020 Vienna, Austria}

\author{R.~B\"ucker}
 \affiliation{Atominstitut, TU-Wien, 1020 Vienna, Austria}

 \author{T.~Berrada}
 \affiliation{Atominstitut, TU-Wien, 1020 Vienna, Austria}
 
  \author{Ch.~Koller }
  \affiliation{Atominstitut, TU-Wien, 1020 Vienna, Austria}
  
\author{G.~Kazakov}  
\affiliation{Wolfgang Pauli Institute, University of Vienna, 1090 Vienna, Austria}
 \affiliation{State Polytechnic University, 195251 St.Petersburg, Russia}

 \author{I.E.~Mazets}
 \affiliation{Atominstitut, TU-Wien, 1020 Vienna, Austria}
 \affiliation{Wolfgang Pauli Institute, University of Vienna, 1090 Vienna, Austria}
 \affiliation{Ioffe Physico-Technical Institute, 194021 St.Petersburg, Russia}

\author{H.-P.~Stimming}
\affiliation{Wolfgang Pauli Institute, University of Vienna, 1090 Vienna, Austria}

\author{A.~Perrin}
\affiliation{Atominstitut, TU-Wien, 1020 Vienna, Austria}
\affiliation{Wolfgang Pauli Institute, University of Vienna, 1090 Vienna, Austria}

\author{T.~Schumm}
\affiliation{Atominstitut, TU-Wien, 1020 Vienna, Austria}
\affiliation{Wolfgang Pauli Institute, University of Vienna, 1090 Vienna, Austria} 

 \author{J.~Schmiedmayer}
  \affiliation{Atominstitut, TU-Wien, 1020 Vienna, Austria}

\title{Two-point phase correlations of a one-dimensional bosonic Josephson junction}

\begin{abstract}
We realize a one-dimensional Josephson junction using quantum degenerate Bose gases in a tunable double well potential on an atom chip. Matter wave interferometry gives direct access to the relative phase field, which reflects the interplay of thermally driven fluctuations and phase locking due to tunneling. The thermal equilibrium state is characterized by probing the full statistical distribution function of the two-point phase correlation. Comparison to a stochastic model allows to measure the coupling strength and temperature and hence a full characterization of the system.
\end{abstract}

\date{\today}
\maketitle

Josephson dynamics between weakly coupled macroscopic wave functions have been observed in superconductors~\cite{jose62, Likharev1979}, superfluid Helium~\cite{Sukhatme2001, Pereverzev1997}, and recently using Bose-Einstein condensates in double well potentials ~\cite{Albiez2005a,Levy2007,Leblanc2010}. The bosonic Josephson junction (BJJ) is especially interesting, as particle interactions lead to additional dynamical modes such as quantum self trapping or $\pi$ phase modes~\cite{ragh99, Albiez2005a} and finite temperature leads to enhanced fluctuations of the observables~\cite{Gati2006}. In contrast to other implementations, the BJJ enables complete experimental control over all relevant system parameters such as the coupling strength or relative population together with direct access to the conjugate observables number and phase. Theoretical work has mostly employed a two-mode approach to describe the finite temperature equilibrium system and dynamical properties~\cite{ragh99, bouc03}.

One-dimensional (1D) Josephson junctions show a significantly enriched physical behavior, as the two involved wave functions can not be described by single quantum modes any more. The non-interacting 1D junction represents an implementation of the Sine-Gordon Hamiltonian which occurs in widespread areas of physics~\cite{Likharev1986a, Ustinov1998}. In the 1D bosonic junction interactions and finite temperature are expected to cause dynamical instabilities of the classical Josephson modes~\cite{Bouchoule2005}. Whether quasi-static phenomena such as quantum self-trapping persist in 1D is issue of ongoing discussion~\cite{Hipolito2010}. 


In this work we realize and fully characterize a one-dimensional bosonic Josephson junction using quantum degenerate Bose gases in a tunable double well potential. The finite temperature equilibrium state is marked by the competing effects of thermally driven phase fluctuations and phase locking due to tunnel coupling. Fluctuations of the relative population are $<1\,\%$ and can be neglected~\cite{Gati2006}. We probe the coherence properties of the coupled system by performing matter wave interferometry. Comparing the statistical distribution function of two-point phase correlations to a stochastic model~\cite{Mauser, bouc03}, we measure the coupling energy or the temperature of the system. 

\begin{figure}
\begin{center}
\includegraphics[]{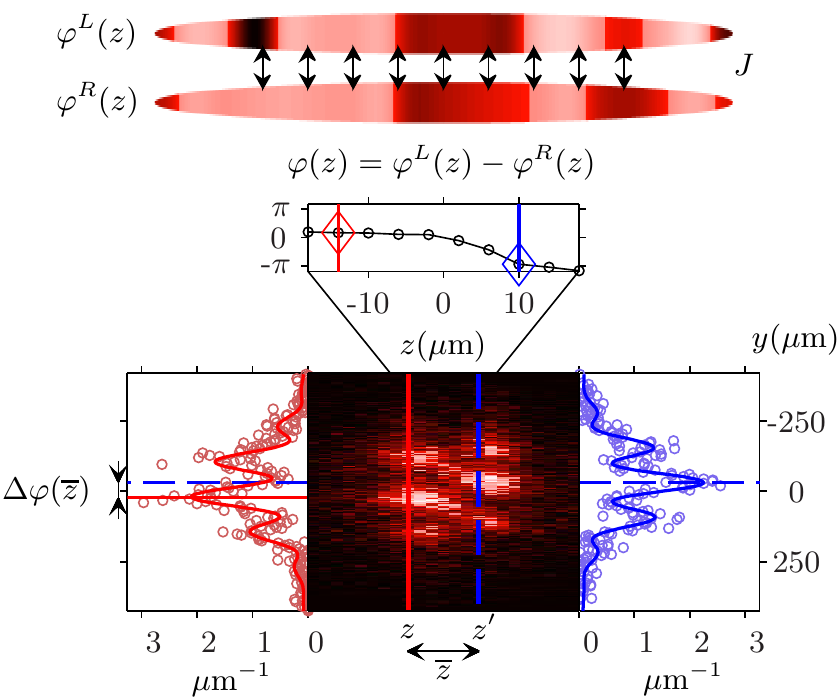}
\caption{Two coupled Bose gases are released from a double well potential. The matter wave interference pattern emerging during expansion gives access to the relative phase $\varphi(z)$ along the samples. Fluctuations in the absolute phase of the two Bose gases transform into visible density fluctuations~\cite{Imam2010, Manz2010}. We characterize two-point phase correlations of the system by measuring the statistical properties of the difference of relative phases $\Delta\varphi(\overline{z})=\varphi(z)-\varphi(z')$.}
\label{fig:experimentalPics}
\end{center}
\end{figure}


The experiments are performed in a horizontally orientated double well potential, generated on an atom chip using radio-frequency (RF) induced adiabatic states~\cite{Schumm2005b, Lesanovsky2006}. Different double well parameters and hence different coupling strengths are realized by using different RF amplitudes, with distances between the minima ranging from 1.2\,$\mu$m to 2\,$\mu$m and a barrier height between $h\times 1.5$\,kHz an $h\times12$\,kHz. The trap frequencies of the individual wells are measured as $\nu_{y}=2.9$~kHz and $\nu_{x}=3.3$~kHz along the tightly confined directions and $\nu_{z}=18 $\,Hz along the longitudinal direction. The two samples are coupled along the strongly confining direction $y$. Note that the double well potentials remains static throughout an experimental sequence.

Starting from a thermal gas of $^{87}$Rb atoms in the $\left|F=1,m_F=-1\right\rangle$ state, we create a system of coupled degenerate Bose gases of adjustable temperature using forced evaporative cooling in the static double well potential. This, together with a phase of plain evaporation of at least 180\,ms, ensures a system in thermal equilibrium. Each well contains 3200(510) atoms, corresponding to an in-situ peak line density of 72(7)\,$\mu$m$^{-1}$, and a chemical potential of $\mu/h=1.72(0.17)$\,kHz~\cite{Gerbier2004}. With temperatures of typically 150\,nK, a regime where $\mu<k_B T\lesssim h\nu_{x,y}$ is realized.  As the physics considered here concerns only long-wavelength excitations, each sample can be regarded as one-dimensional ~\cite{Petrov2000c, Gerbier2004}~\footnote{The fraction of atoms  in transversally excited states is estimated to 5\% (150\,nK), it is assumed to have no influence on the correlation properties of the condensed part of the sample.}.


To probe phase correlations of the BJJ we perform matter wave interferometry. After suddenly ($<10\,\mu$s) switching off all confining potentials the two samples overlap in 46\,ms time-of-flight (TOF) expansion. The emerging interference pattern is recorded using a light sheet fluorescence imaging~\cite{Bucker2009}. The relative phase $\varphi(z)=\varphi^L(z)-\varphi^R(z)$ between the left ($L$) and right ($R$) Bose gas is determined by fitting a cosine function with a gaussian envelope to the density profile in each of the $4\,\mu$m wide pixel slices (see Fig.~\ref{fig:experimentalPics}). The fitting introduces an uncertainty $\delta\varphi(z)$ on the value of $\varphi(z)$ which follows a normal distribution with zero mean and a standard deviation ranging from 0.1 to 0.5 radians. The analysis is restricted to a maximum distance of 40\,$\mu$m over which the atomic density drops by $30\,\%$. The experiment is repeated 500 times for each parameter set to provide a full analysis of the statistical properties of $\varphi(z)$.


A single trapped 1D Bose gas at finite temperature is characterized by a density profile $\rho_{0}(z)$ and a spatially fluctuating phase~\footnote{Quantum fluctuations can be neglected in the temperature regime considered here.} responsible for the exponential decay of the first order correlation function on the length scale $\lambda_T=2\hbar^2\rho_{0}/(m k_B T)$~\cite{Petrov2000c}. Interference of two \emph{independent} samples can be used to probe the coherence properties of \emph{single} 1D Bose gases~\cite{Hofferberth2008, Gritsev2006, Hofferberth2007c}. 

Introducing a tunnel coupling strength $J$ between two 1D Bose gases implements a \emph{combined} system with an additional degree of freedom in the \emph{relative} phase $\varphi(z)$~\cite{bouc03}\footnote{Note that $\lambda_T$ is independent of $J$.}. It is associated with the additional length scale $l_J = \sqrt{\hbar/4m J}$, which represents the typical distance on which tunnel coupling restores a spatially constant relative phase. For $l_J<\lambda_T$ the system becomes phase-locked, i.e. the coupling counterbalances the randomization due to thermal excitations.

\begin{figure}
\begin{center}
\includegraphics[scale=1]{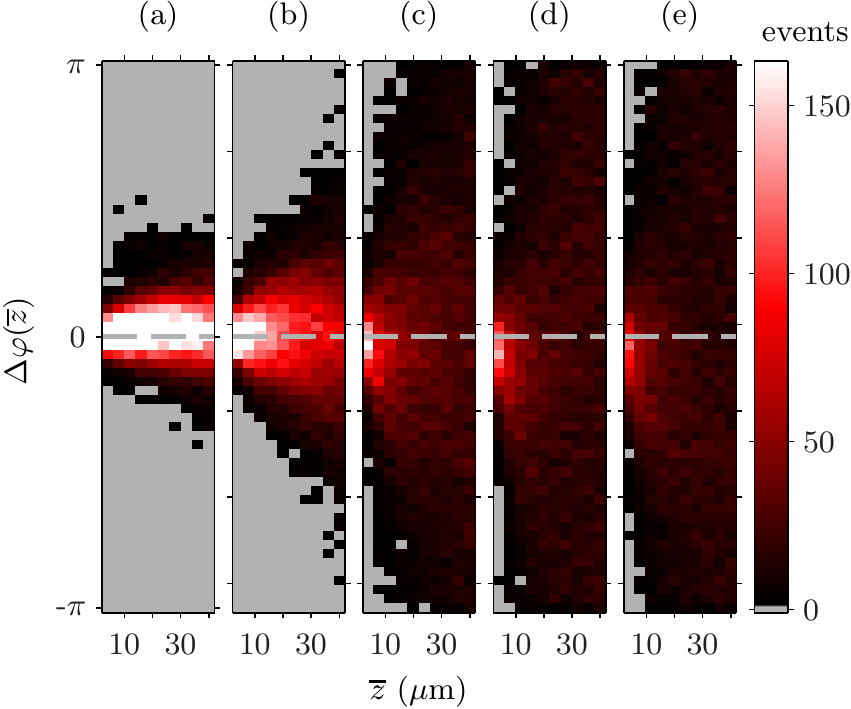}
\caption{Distributions of the difference of the relative phase $\Delta\varphi(\overline{z})$ for decreasing tunnel coupling (from left to right, compare table~\ref{tab:compareCoupling}). (a,b) High coupling yields a narrow distribution and high coherence over the entire length of the system. (c-e) Decreasing coupling leads to an increasingly fast loss of spatial phase correlations and thus a randomization of relative phases. A slight bias of the distribution can be assigned to a small relative velocity (20\,$\mu$m/s) of the two samples during the expansion. }
\label{fig:PhaseDensities}
\end{center}
\end{figure}

Two-point phase correlations along the coupled 1D samples can be probed through the statistical properties of the difference of the relative phase $\Delta\varphi(\overline{z})=\varphi(z)-\varphi(z')$ where $\overline{z}=\left|z-z'\right|$. The scaling of the distribution of $\Delta\varphi(\overline{z})$ with $\overline{z}$ gives access to the spatial extension of the relative phase correlations and hence to the relevant length scales $\lambda_T$ and $l_J$~\cite{bouc03}. A narrow distribution of $\Delta\varphi(\overline{z})$ (peaked around 0) indicates high coherence and phase locking, whereas a broad distribution (between $-\pi$ and $\pi$) characterizes uncorrelated phases. Figure~\ref{fig:PhaseDensities} depicts measured distributions of $\Delta\varphi(\overline{z})$ for different distances and tunnel couplings, ranging from strongly phase locked (a) to almost independent samples (e). 

\begin{figure}
\begin{center}
\includegraphics{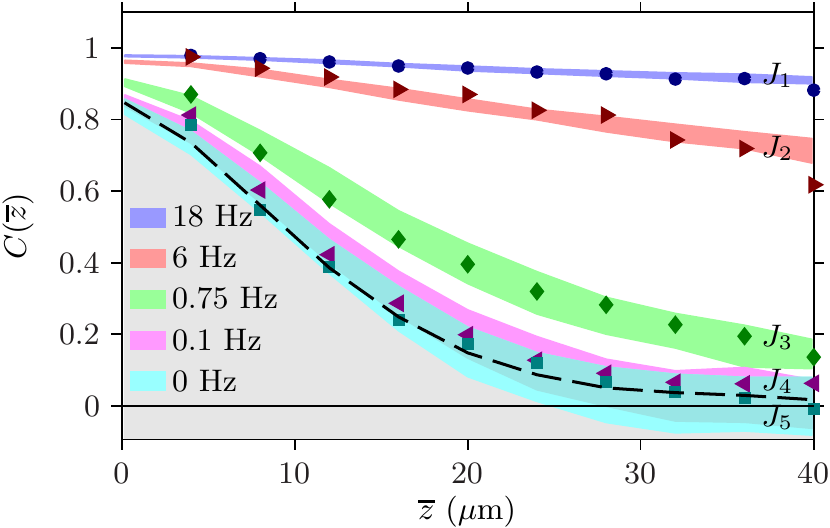}
\caption{Real part of the phase correlation function $C(\overline{z})$ for different couplings and $T\simeq 155$~nK (compare Table~\ref{tab:compareCoupling}). The symbols represent experimental values, derived from the data shown in Fig.~\ref{fig:PhaseDensities}. A stochastic OU model allows to estimate the couplings $J_1$ to $J_5$ (compare Table~\ref{tab:compareCoupling}). The colored areas depict two standard deviations on the values of $C(\overline{z})$ obtained by repeating the analysis 30 times. A simulation of $C(\overline{z})$ for completely uncoupled Bose gases is displayed as a black dashed line and indicates the limits of our imaging resolution.}
\label{fig:CorrelFig}
\end{center}
\end{figure}

\begin{figure}
\begin{center}
\includegraphics{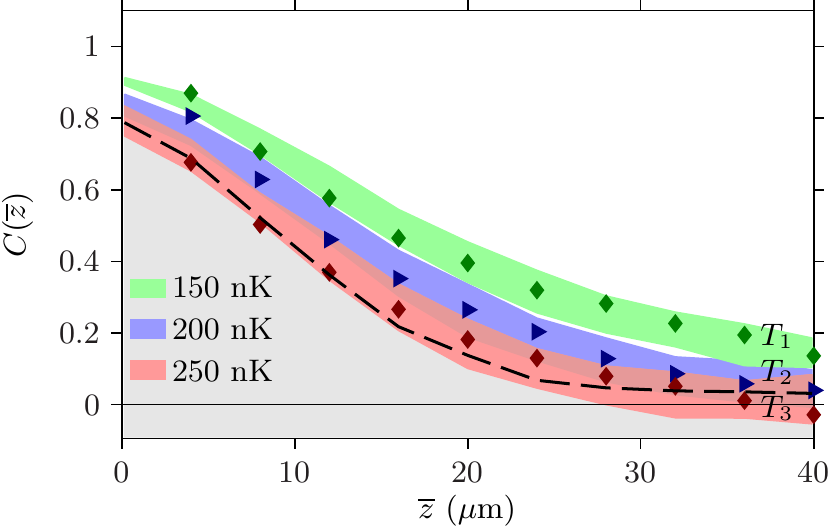}
\caption{Real part of the phase correlation function $C(\overline{z})$ for fixed coupling $J_3$ and different temperatures $T_1$ to $T_3$. Comparison with the stochastic simulation allows thermometry of the 1D bosonic Josephson junction (compare Table~\ref{tab:compareCoupling}).}
\label{fig:CorrelFig_DiffT}
\end{center}
\end{figure}

Calculating the real part of the \emph{phase correlation function} $C(\overline{z})=\text{Re}\left\langle e^{i\varphi(z)-i\varphi(z')}\right\rangle=\left\langle\cos(\Delta\varphi(\overline{z}))\right\rangle$ allows to quantify the spread of the distribution of $\Delta\varphi(\overline{z})$~\cite{Mauser}. By changing from 1 to 0, $C(\overline{z})$ describes the transition from spatially locked to uncorrelated relative phases for increasing $\overline{z}$. The characteristic length scale and shape of this decay is determined by the BJJ parameters $\rho_0$, $J$, and $T$. Figures~\ref{fig:CorrelFig} and~\ref{fig:CorrelFig_DiffT} show measured values of $C(\overline{z})$ for different values of $J$ and $T$ at constant atom number.


To compare our results to theoretical predictions of~\cite{bouc03} we simulate single realizations of $\varphi(z)$ using an Ornstein-Uhlenbeck (OU) stochastic process~\cite{Mauser}. These realizations exactly follow the phase correlations predicted in~\cite{bouc03} for uniform systems but allow us to account for a finite number of experimental realizations and finite system size along with a spatially varying density $\rho_0(z)$ (calculated according to~\cite{Gerbier2004}) and a density-dependent coupling $J(\rho_0(z))$ within a local density approach~\cite{EPAPS}. From this we construct single realizations of the in-situ wave functions $\psi_0^{L,R}(z)=\sqrt{\rho_0(z)}\exp\left(i\varphi_0^{L,R}(z)\right)$ each representing single outcomes of the experiment.


When comparing to experimental data, we have to include the effects of the expansion and the detection process. Since the expansion of the system can be considered ballistic~\cite{Imam2010, Manz2010} we numerically compute $\left|\psi^{L,R}\right\rangle=\hat{U}(t)\left|\psi_0^{L,R}\right\rangle$ for a TOF $t=46$~ms, where $\hat{U}(t)$ is the free evolution operator. The resulting relative phase $\varphi(z)=\varphi^L(z)-\varphi^R(z)$ and the density $\rho(z)$ after expansion are then individually convoluted with the point-spread-function of the imaging system, which we estimate to be gaussian with an RMS width of 7\,$\mu$m~\cite{Bucker2009}. Finally the uncertainty $\delta\varphi(z)$ introduced by the fitting procedure is added to the resulting relative phase, coarse grained to the 4\,$\mu$m pixel size. 

Averaging over 500 realizations of the process as in the experiment, the experimental results can be reproduced by adjusting the simulation parameter $\rho_0$, $J_{\rm OU}$, and $T_{\rm OU}$. Repeating the described procedure many times allows to evaluate the uncertainty on the estimation of $C(\overline{z})$ due to the finite number of experimental realizations.


Figure~\ref{fig:CorrelFig} compares measured and simulated values of $C(\overline{z})$ for five different tunnel couplings  ($J_1$ to $J_5$), with fixed temperature $T_{\rm TOF}$ and density profile $\rho_0(z)$. The temperature $T_{\rm TOF}$ is determined by fitting a Bose function to the thermal wings of the recorded density profiles~\cite{Bucker2009}. To match the OU simulations of $C(\overline{z})$ with the experimental data, only the parameters $J_1$ to $J_5$ are adjusted (see Fig.~\ref{fig:CorrelFig} and Table~\ref{tab:compareCoupling}). 

This correlation analysis provides a direct experimental measure of the coupling strength in a BJJ, without the necessity to probe dynamical properties or rely on simulations of the system in the double well potential. We show here that our experimental control allows to adjust the tunnel coupling over two orders of magnitudes. 

A comparison with an alternative numerical simulation based on the time-dependent 1D Gross-Pitaevskii equation shows, within a factor of two, a good agreement with our results (compare Table~\ref{tab:compareCoupling} and see~\cite{EPAPS}). Note that a factor of two in tunnel coupling corresponds to less than 3\,$\%$ uncertainty in parameters defining the double well potential, highlighting the need for a direct and precise experimental measurement of $J$, as realized in this work.

In cases where the tunnel coupling is known, the analysis of $C(\overline{z})$ can be used for thermometry of the Josephson junction as shown in Fig.~\ref{fig:CorrelFig_DiffT} ~\cite{Gati2006}. Here, the trapping potential remains unchanged and $J$ is fixed to $J_3$. The temperature $T$ is set by changing the final position of the cooling RF-field while keeping the total atom number constant.  The temperatures $T_{\rm OU}$ obtained by adjusting the OU simulations to the data are in excellent agreement with the independently measured $T_{\rm TOF}$ (compare Table~\ref{tab:compareCoupling}). For experimental systems featuring an optical resolution better than the thermal coherence length, the analysis presented in this work would even allow the simultaneous determination of the coupling $J$ and the temperature $T$. 

\begin{table}[hbt]
	\centering
		\begin{tabular}{|c||c|c|c|c|c|c|}
			\hline
			 Fig.~\ref{fig:CorrelFig}&$T_{\rm TOF}$(nK)&$J_{\rm OU}$(Hz)&$J_{\rm GP}$(Hz)&$\lambda_T(\mu m)$&$l_J(\mu m)$\\\hline\hline
			$J_1$&154(5)&16.5-21 & 35&5.8&2.9-3.3\\\hline
			$J_2$&150(5)&5-7.5&4.8 &5.4&4.9-6\\\hline
			$J_3$&153(5)&0.65-0.9&0.15&5.3&14.4-16.8\\\hline
			$J_4$&154(5)&0.05-0.15& 0.05&5.1&34.9-60.4\\\hline
			$J_5$&163(5)& $<$0.1& $<$0.05&4.8&$>$47.8\\\hline
			\hline 
			 Fig.~\ref{fig:CorrelFig_DiffT}&$T_{\rm TOF}$(nK)&$T_{\rm OU}$(nK)&$J_3 $(Hz)&$\lambda_T(\mu m)$&$l_J(\mu m)$\\\hline\hline
			$T_1$&155(10)&125-180&0.75&4.5-6.5 &15.6\\\hline
		  $T_2$&210(5)&180-230&0.75 &2.5-3.2&15.6\\\hline
			$T_3$&275(5)&240-280&0.75 &1.5-1.8&15.6\\\hline
		\end{tabular}		
		\caption{Comparison of experimentally measured and simulated parameters. All spatially dependent values are given at the position of the peak atomic density ($z=0$).}
		\label{tab:compareCoupling}
\end{table}

\begin{figure}
\begin{center}
\includegraphics{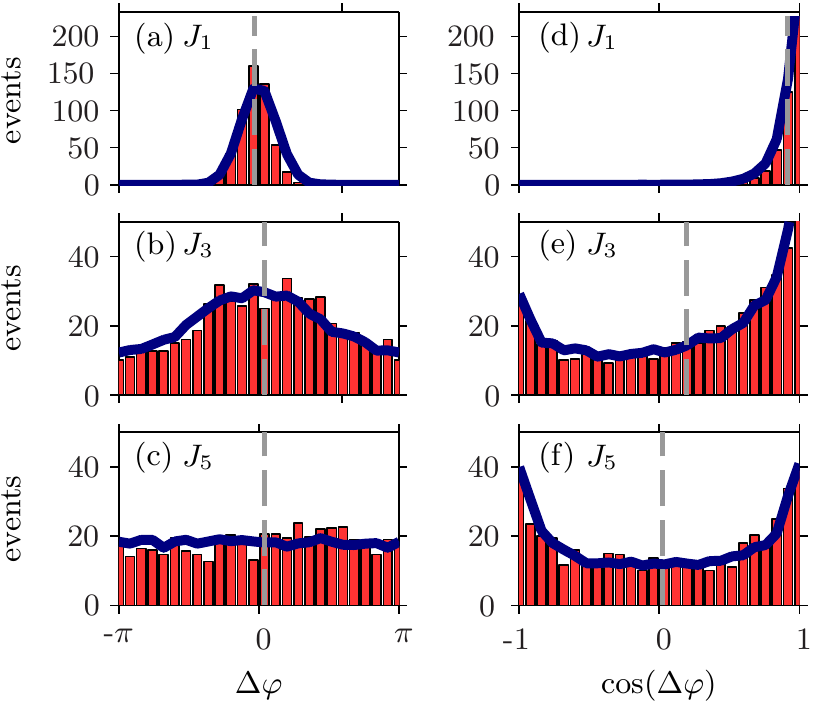}
\caption{Distributions of $\Delta\varphi(\overline{z})$ (a-c) and $\cos(\Delta\varphi(\overline{z}))$ (d-f) for $\overline{z}=32\,\mu$m and  three different couplings ($J_1$, $J_3$, $J_5$). Red histograms represent measured, blue lines simulated distributions for 500 realizations. (a-c) The mean value of $\Delta\varphi(\overline{z})$ (dashed gray line) does not change while the phase spreads. (d-f) The mean value of $\cos\Delta\varphi(\overline{z})$ (dashed gray line) decays from  about $1$ (strongly coupled) to 0 (uncoupled).}
\label{fig:PhaseAndCosPhaseDistribution}
\end{center}
\end{figure}

The experimental data as well as the OU simulation give access not only to expectation values but also to the full distribution of $\Delta\varphi(\overline{z})$. Figure~\ref{fig:PhaseAndCosPhaseDistribution} shows exemplary distributions of $\Delta\varphi(\overline{z})$ and its cosine, for $\overline{z}=32\,\mu$m and couplings $J_1$, $J_3$, and $J_5$. The excellent agreement between experiment and simulation indicates that we have indeed realized an equilibrium system as decribed in~\cite{bouc03}.


In summary we have used two-point phase correlations to fully characterize a tunable 1D bosonic Josephson junction in thermal equilibrium. Comparing the experimental data to the results of a stochastic model based on an Ornstein-Uhlenbeck process allows us to determine the coupling strength or the temperature of the system. This full characterization marks the starting point for further studies of dynamical properties and non-equilibrium states of coupled one-dimensional systems. Further research will investigate the interplay of spatial density and phase correlations.

We acknowlege support from the FWF projects P21080-N16 and P22590-N16, the WPI Thematic Program QUANTUM-10 and the EU projects MIDAS and Marie Curie (FP7 GA $\rm n^o$~236702).


\begin{thebibliography}{29}
\expandafter\ifx\csname natexlab\endcsname\relax\def\natexlab#1{#1}\fi
\expandafter\ifx\csname bibnamefont\endcsname\relax
  \def\bibnamefont#1{#1}\fi
\expandafter\ifx\csname bibfnamefont\endcsname\relax
  \def\bibfnamefont#1{#1}\fi
\expandafter\ifx\csname citenamefont\endcsname\relax
  \def\citenamefont#1{#1}\fi
\expandafter\ifx\csname url\endcsname\relax
  \def\url#1{\texttt{#1}}\fi
\expandafter\ifx\csname urlprefix\endcsname\relax\def\urlprefix{URL }\fi
\providecommand{\bibinfo}[2]{#2}
\providecommand{\eprint}[2][]{\url{#2}}

\bibitem[{\citenamefont{Josephson}(1962)}]{jose62}
\bibinfo{author}{\bibfnamefont{B.~D.} \bibnamefont{Josephson}},
  \bibinfo{journal}{Phys. Lett.} \textbf{\bibinfo{volume}{1}},
  \bibinfo{pages}{251 } (\bibinfo{year}{1962}).

\bibitem[{\citenamefont{Likharev}(1979)}]{Likharev1979}
\bibinfo{author}{\bibfnamefont{K.~K.} \bibnamefont{Likharev}},
  \bibinfo{journal}{Rev. Mod. Phys.} \textbf{\bibinfo{volume}{51}},
  \bibinfo{pages}{101} (\bibinfo{year}{1979}).

\bibitem[{\citenamefont{Sukhatme et~al.}(2001)\citenamefont{Sukhatme,
  Mukharsky, Chui, and Pearson}}]{Sukhatme2001}
\bibinfo{author}{\bibfnamefont{K.}~\bibnamefont{Sukhatme}},
  \bibinfo{author}{\bibfnamefont{Y.}~\bibnamefont{Mukharsky}},
  \bibinfo{author}{\bibfnamefont{T.}~\bibnamefont{Chui}}, \bibnamefont{and}
  \bibinfo{author}{\bibfnamefont{D.}~\bibnamefont{Pearson}},
  \bibinfo{journal}{Nature (London)} \textbf{\bibinfo{volume}{411}},
  \bibinfo{pages}{17} (\bibinfo{year}{2001}).

\bibitem[{\citenamefont{Pereverzev et~al.}(1997)\citenamefont{Pereverzev,
  Loshak, Backhaus, and Davis}}]{Pereverzev1997}
\bibinfo{author}{\bibfnamefont{S.~V.} \bibnamefont{Pereverzev}},
  \bibinfo{author}{\bibfnamefont{A.}~\bibnamefont{Loshak}},
  \bibinfo{author}{\bibfnamefont{S.}~\bibnamefont{Backhaus}}, \bibnamefont{and}
  \bibinfo{author}{\bibfnamefont{J.~C.} \bibnamefont{Davis}},
  \bibinfo{journal}{Nature (London)} \textbf{\bibinfo{volume}{388}},
  \bibinfo{pages}{449} (\bibinfo{year}{1997}).

\bibitem[{\citenamefont{Albiez et~al.}(2005)\citenamefont{Albiez, Gati,
  F\"olling, Hunsmann, Cristiani, and Oberthaler}}]{Albiez2005a}
\bibinfo{author}{\bibfnamefont{M.}~\bibnamefont{Albiez \emph{et al.}}}, \bibinfo{journal}{Phys. Rev. Lett.}
  \textbf{\bibinfo{volume}{95}}, \bibinfo{pages}{010402}
  (\bibinfo{year}{2005}).

\bibitem[{\citenamefont{Levy et~al.}(2007)\citenamefont{Levy, Lahoud, Shomroni,
  and Steinhauer}}]{Levy2007}
\bibinfo{author}{\bibfnamefont{S.}~\bibnamefont{Levy}},
  \bibinfo{author}{\bibfnamefont{E.}~\bibnamefont{Lahoud}},
  \bibinfo{author}{\bibfnamefont{I.}~\bibnamefont{Shomroni}}, \bibnamefont{and}
  \bibinfo{author}{\bibfnamefont{J.}~\bibnamefont{Steinhauer}},
  \bibinfo{journal}{Nature (London)} \textbf{\bibinfo{volume}{449}},
  \bibinfo{pages}{579} (\bibinfo{year}{2007}).

\bibitem[{\citenamefont{Leblanc et~al.}(2010)\citenamefont{Leblanc, Bardon,
  Mckeever, Extavour, Jervis, Thywissen, Piazza, and Smerzi}}]{Leblanc2010}
\bibinfo{author}{\bibfnamefont{L.~J.} \bibnamefont{Leblanc \emph{et al.}}}
  (\bibinfo{year}{2010}), \eprint{arXiv:1006.3550v1}.

\bibitem[{\citenamefont{Raghavan et~al.}(1999)\citenamefont{Raghavan, Smerzi,
  Fantoni, and Shenoy}}]{ragh99}
\bibinfo{author}{\bibfnamefont{S.}~\bibnamefont{Raghavan}},
  \bibinfo{author}{\bibfnamefont{A.}~\bibnamefont{Smerzi}},
  \bibinfo{author}{\bibfnamefont{S.}~\bibnamefont{Fantoni}}, \bibnamefont{and}
  \bibinfo{author}{\bibfnamefont{S.~R.} \bibnamefont{Shenoy}},
  \bibinfo{journal}{Phys. Rev. A} \textbf{\bibinfo{volume}{59}},
  \bibinfo{pages}{620} (\bibinfo{year}{1999}).

\bibitem[{\citenamefont{Gati et~al.}(2006)\citenamefont{Gati, Hemmerling,
  F\"olling, Albiez, and Oberthaler}}]{Gati2006}
\bibinfo{author}{\bibfnamefont{R.}~\bibnamefont{Gati \emph{et al.}}},
  \bibinfo{journal}{Phys. Rev. Lett.} \textbf{\bibinfo{volume}{96}},
  \bibinfo{pages}{130404} (\bibinfo{year}{2006}).

\bibitem[{\citenamefont{Whitlock and Bouchoule}(2003)}]{bouc03}
\bibinfo{author}{\bibfnamefont{N.~K.} \bibnamefont{Whitlock}} \bibnamefont{and}
  \bibinfo{author}{\bibfnamefont{I.}~\bibnamefont{Bouchoule}},
  \bibinfo{journal}{Phys. Rev. A} \textbf{\bibinfo{volume}{68}},
  \bibinfo{pages}{053609} (\bibinfo{year}{2003}).

\bibitem[{\citenamefont{Likharev}(1986)}]{Likharev1986a}
\bibinfo{author}{\bibfnamefont{K.}~\bibnamefont{Likharev}},
  \emph{\bibinfo{title}{{Dynamics of Josephson Junctions and Circuits}}}
  (\bibinfo{publisher}{Gordon and Breach Science Publishers},
  \bibinfo{address}{New York}, \bibinfo{year}{1986}).

\bibitem[{\citenamefont{Ustinov}(1998)}]{Ustinov1998}
\bibinfo{author}{\bibfnamefont{A.}~\bibnamefont{Ustinov}},
  \bibinfo{journal}{Physica D} \textbf{\bibinfo{volume}{123}},
  \bibinfo{pages}{315} (\bibinfo{year}{1998}).

\bibitem[{\citenamefont{Bouchoule}(2005)}]{Bouchoule2005}
\bibinfo{author}{\bibfnamefont{I.}~\bibnamefont{Bouchoule}},
  \bibinfo{journal}{Eur. Phys. J. D} \textbf{\bibinfo{volume}{35}},
  \bibinfo{pages}{147} (\bibinfo{year}{2005}).

\bibitem[{\citenamefont{Hipolito and Polkovnikov}(2010)}]{Hipolito2010}
\bibinfo{author}{\bibfnamefont{R.}~\bibnamefont{Hipolito}} \bibnamefont{and}
  \bibinfo{author}{\bibfnamefont{A.}~\bibnamefont{Polkovnikov}},
  \bibinfo{journal}{Phys. Rev. A} \textbf{\bibinfo{volume}{81}},
  \bibinfo{pages}{013621} (\bibinfo{year}{2010}).

\bibitem[{\citenamefont{Stimming et~al.}(2010)\citenamefont{Stimming, Mauser,
  Schmiedmayer, and Mazets}}]{Mauser}
\bibinfo{author}{\bibfnamefont{H.~P.} \bibnamefont{Stimming}},
  \bibinfo{author}{\bibfnamefont{N.~J.} \bibnamefont{Mauser}},
  \bibinfo{author}{\bibfnamefont{J.}~\bibnamefont{Schmiedmayer}},
  \bibnamefont{and} \bibinfo{author}{\bibfnamefont{I.~E.}
  \bibnamefont{Mazets}}, \bibinfo{journal}{Phys. Rev. Lett.}
  \textbf{\bibinfo{volume}{105}}, \bibinfo{pages}{015301}
  (\bibinfo{year}{2010}).

\bibitem[{\citenamefont{Imambekov et~al.}(2009)\citenamefont{Imambekov, Mazets,
  Petrov, Gritsev, Manz, Hofferberth, Schumm, Demler, and
  Schmiedmayer}}]{Imam2010}
\bibinfo{author}{\bibfnamefont{A.}~\bibnamefont{Imambekov \emph{et al.}}},
  \bibinfo{journal}{Phys. Rev. A} \textbf{\bibinfo{volume}{80}},
  \bibinfo{pages}{033604} (\bibinfo{year}{2009}).

\bibitem[{\citenamefont{Manz et~al.}(2010)\citenamefont{Manz, B\"ucker, Betz,
  Koller, Hofferberth, Mazets, Imambekov, Demler, Perrin, Schmiedmayer
  et~al.}}]{Manz2010}
\bibinfo{author}{\bibfnamefont{S.}~\bibnamefont{Manz \emph{et al.}}},
  \bibnamefont{et~al.}, \bibinfo{journal}{Phys. Rev. A}
  \textbf{\bibinfo{volume}{81}}, \bibinfo{pages}{031610}
  (\bibinfo{year}{2010}).

\bibitem[{\citenamefont{Schumm et~al.}(2005)\citenamefont{Schumm, Hofferberth,
  Andersson, Wildermuth, Groth, Bar-Joseph, Schmiedmayer, and
  Kr\"{u}ger}}]{Schumm2005b}
\bibinfo{author}{\bibfnamefont{T.}~\bibnamefont{Schumm \emph{et al.}}},
  \bibinfo{journal}{Nature Physics} \textbf{\bibinfo{volume}{1}},
  \bibinfo{pages}{57} (\bibinfo{year}{2005}).

\bibitem[{\citenamefont{Lesanovsky et~al.}(2006)\citenamefont{Lesanovsky,
  Schumm, Hofferberth, Andersson, Kr\"{u}ger, and
  Schmiedmayer}}]{Lesanovsky2006}
\bibinfo{author}{\bibfnamefont{I.}~\bibnamefont{Lesanovsky \emph{et al.}}},
  \bibinfo{journal}{Phys. Rev. A} \textbf{\bibinfo{volume}{73}},
  \bibinfo{pages}{033619} (\bibinfo{year}{2006}).

\bibitem[{\citenamefont{Gerbier}(2004)}]{Gerbier2004}
\bibinfo{author}{\bibfnamefont{F.}~\bibnamefont{Gerbier}},
  \bibinfo{journal}{EPL (Europhysics Letters)} \textbf{\bibinfo{volume}{66}},
  \bibinfo{pages}{771} (\bibinfo{year}{2004}).

\bibitem[{\citenamefont{Petrov et~al.}(2000)\citenamefont{Petrov, Shlyapnikov,
  and Walraven}}]{Petrov2000c}
\bibinfo{author}{\bibfnamefont{D.~S.} \bibnamefont{Petrov}},
  \bibinfo{author}{\bibfnamefont{G.~V.} \bibnamefont{Shlyapnikov}},
  \bibnamefont{and} \bibinfo{author}{\bibfnamefont{J.~T.~M.}
  \bibnamefont{Walraven}}, \bibinfo{journal}{Phys. Rev. Lett.}
  \textbf{\bibinfo{volume}{85}}, \bibinfo{pages}{3745} (\bibinfo{year}{2000}).

\bibitem[{Note1()}]{Note1}
 \bibinfo{note}{The fraction of atoms  in transversally excited states is estimated to 5\% (150\,nK), it is assumed to have no influence on the correlation properties of the condensed part of the sample.}

\bibitem[{\citenamefont{B\"{u}cker et~al.}(2009)\citenamefont{B\"{u}cker,
  Perrin, Manz, Betz, Koller, Plisson, Rottmann, Schumm, and
  Schmiedmayer}}]{Bucker2009}
\bibinfo{author}{\bibfnamefont{R.}~\bibnamefont{B\"{u}cker \emph{et al.}}},
  \bibinfo{journal}{New J. Phys.} \textbf{\bibinfo{volume}{11}},
  \bibinfo{pages}{103039} (\bibinfo{year}{2009}).

\bibitem[{Note2()}]{Note2}
 \bibinfo{note}{Quantum fluctuations can be neglected in the temperature
  regime considered here.}

\bibitem[{\citenamefont{Hofferberth et~al.}(2008)\citenamefont{Hofferberth,
  Lesanovsky, Schumm, Imambekov, Gritsev, Demler, and
  Schmiedmayer}}]{Hofferberth2008}
\bibinfo{author}{\bibfnamefont{S.}~\bibnamefont{Hofferberth \emph{et al.}}},
  \bibinfo{journal}{Nature Physics} \textbf{\bibinfo{volume}{4}},
  \bibinfo{pages}{489} (\bibinfo{year}{2008}).

\bibitem[{\citenamefont{Gritsev et~al.}(2006)\citenamefont{Gritsev, Altman,
  Demler, and Polkovnikov}}]{Gritsev2006}
\bibinfo{author}{\bibfnamefont{V.}~\bibnamefont{Gritsev}},
  \bibinfo{author}{\bibfnamefont{E.}~\bibnamefont{Altman}},
  \bibinfo{author}{\bibfnamefont{E.}~\bibnamefont{Demler}}, \bibnamefont{and}
  \bibinfo{author}{\bibfnamefont{A.}~\bibnamefont{Polkovnikov}},
  \bibinfo{journal}{Nature Physics} \textbf{\bibinfo{volume}{2}},
  \bibinfo{pages}{705} (\bibinfo{year}{2006}).

\bibitem[{\citenamefont{Hofferberth et~al.}(2007)\citenamefont{Hofferberth,
  Lesanovsky, Fischer, Schumm, and Schmiedmayer}}]{Hofferberth2007c}
\bibinfo{author}{\bibfnamefont{S.}~\bibnamefont{Hofferberth \emph{et al.}}},
  \bibinfo{journal}{Nature (London)} \textbf{\bibinfo{volume}{449}},
  \bibinfo{pages}{324} (\bibinfo{year}{2007}).

\bibitem[{Note3()}]{Note3}
 \bibinfo{note}{Note that $\lambda _T$ is independent of $J$.}

\bibitem[{EPA()}]{EPAPS}
\bibinfo{note}{See EPAPS Document No. X for supplementary information. For more
  information on EPAPS, see http://www.aip.org/pubservs/epaps.html.}

\end{thebibliography}
\end{document}